\documentclass[aps,prl,twocolumn,showpacs,preprintnumbers,a4paper]{revtex4} 

\usepackage{graphics}
\usepackage{epsf}
\usepackage{epsfig}

\begin{document} 


\title{Universal scaling behavior at the upper critical dimension\\
of non-equilibrium continuous phase transitions}

\author{S. L\"ubeck}
\affiliation{Weizmann Institute of Science, 
Department of Physics of Complex Systems, 
76100 Rehovot, Israel,
}

\author{P.\,C.~Heger}
\affiliation{
Institut f\"ur Theoretische Physik,
Universit\"at Duisburg-Essen, 
47048 Duisburg, Germany}

\date{Received 12\,December, 2002, published 12\,June 2003}

\begin{abstract}
In this work we analyze the universal scaling functions and 
the critical exponents at the upper critical dimension
of a continuous phase transition.
The consideration of the universal scaling
behavior yields a decisive check of the value
of the upper critical dimension.
We apply our method to a non-equilibrium
continuous phase transition.
But focusing on the equation of state of the phase transition
it is easy to extend our analysis to all equilibrium
and non-equilibrium phase transitions observed 
numerically or experimentally.
\end{abstract}

\pacs{05.70.Ln, 05.50.+q, 05.65.+b}

\keywords{Phase transition, Universal scaling, upper critical dimension}

\preprint{{\it Physical Review Letters} {\bf 90}, 230601 (2003)}

\maketitle


One of the most impressive features of continuous
phase transitions is the concept of universality
that allows to group the great variety of different types
of critical phenomena into a small number of universality
classes~(see~\cite{STANLEY_1} for a recent review).
All systems belonging to a given universality class
have the same critical exponents and the corresponding 
scaling functions (equation of state, correlation 
functions, etc.) become identical near the critical point.
Classical examples of such universal behavior are for instance 
the coexistence curve of liquid-vapor systems~\cite{GUGGENHEIM_1} 
and the equation of state in ferromagnetic 
systems~(see for instance~\cite{STANLEY_1,MILOSEVIC_2}).
Checking the universality class it is often a more
exacting test to consider scaling functions and 
amplitude combinations (which are just particular values
of the scaling functions) rather than the values of the 
critical exponents.
While for the latter ones the variations between 
different universal classes are often small the amplitude
combinations and therefore the scaling functions may differ
significantly (see~\cite{PRIVMAN_2}).
A foundation for the understanding of the concept of universality 
as well as a tool to estimate the values of the critical exponents
was provided by Wilson's renormalization group (RG) 
approach~\cite{WILSON_1,WILSON_2} 
which maps the critical point onto a fixed point of a
certain transformation of the system's Hamiltonian,
Langevin equation, etc.

Furthermore the RG explains the existence of an 
upper critical dimension~$D_{\text c}$ above which the 
mean-field theory applies
whereas it fails below $D_{\text c}$.
At the upper critical dimension the RG equations yield
mean-field exponents with logarithmic corrections~\cite{WILSON_3}.
These logarithmic corrections make the data analysis
quite difficult and thus most investigations are focused
on the determination of the correction exponents 
(see Eqs.\,(\ref{eq:uni_scal_OPzf},\ref{eq:uni_scal_OPcp}) below) only, lacking the
determination of the scaling functions.

In this work we investigate the universal scaling
behavior of a continuous phase transition at 
$D_{\text c}$ and develop a method of analysis 
that allows us to determine the exponents as well as
the scaling functions.
Therefore we consider three different non-equilibrium
systems exhibiting a continuous phase transition into
an absorbing phase.
Focusing on the equation of state our method can be easily
applied to all equilibrium as well as non-equilibrium
continuous phase transitions observed in numerical simulations
or experiments (as long as the conjugated field can be
physically realized).
In all three models the dynamics obey particle conservation
and according to the universality hypothesis of~\cite{ROSSI_1}
all models are expected to belong to the universality class
of absorbing phase transitions with a conserved field.

The first considered model is the conserved lattice gas (CLG) 
which
was introduced in~\cite{ROSSI_1}.
In the CLG lattice sites may be empty or occupied
by one particle.
In order to mimic a repulsive interaction a given particle
is considered as active if at least one of its
neighboring sites on the lattice is occupied by another
particle.
If all neighboring sites are empty the particle remains
inactive.
Active particles are moved in the next update
step to one of their empty nearest neighbor sites,
selected at random.

The second model is the so-called conserved transfer
threshold process (CTTP)~\cite{ROSSI_1}.
Here, lattice sites may be empty, occupied by one particle,
or occupied by two particles.
Empty and single occupied sites are considered as
inactive whereas double occupied lattice sites are 
considered as active.
In the latter case one tries to transfer both particles
of a given active site to randomly chosen empty or single
occupied nearest neighbor sites.

The third model is a modified version of the
Manna sandpile model~\cite{MANNA_2} the so-called fixed-energy
Manna model~\cite{VESPIGNANI_4}.
In contrast to the CTTP the Manna model allows unlimited
particle occupation of lattice sites.
All lattice sites which are occupied by at least two particles
are considered as active and all 
particles are moved to the neighboring
sites selected at random.

In our simulations (see~\cite{LUEB_22,LUEB_24} for 
details) we start from a random distribution 
of particles and all models reach after a transient 
regime a steady state which is
characterized by the density of active sites~$\rho_{\scriptscriptstyle \text a}$.
The density~$\rho_{\scriptscriptstyle \text a}$ is the order parameter
and the particle density~$\rho$ is the control parameter
of the absorbing phase transition, i.e., the order parameter
vanishes at the critical density~$\rho_{\text c}$ according to
$\rho_{\scriptscriptstyle \text a} \propto \delta\rho^{\beta}$ ,
with the reduced control parameter 
$\delta\rho=\rho/\rho_{\text c}-1$. 
Additionally to the order parameter we consider its 
fluctuations $\Delta \rho_{\scriptscriptstyle \text a}$.
Approaching the transition point from above ($\delta\rho>0$) 
the fluctuations diverge according to 
$\Delta \rho_{\scriptscriptstyle \text a}\propto  \delta\rho^{-\gamma^{\prime}}$
(see~\cite{LUEB_22,LUEB_24}).
Below the critical density (in the absorbing state)
the order parameter as well as its fluctuations
are zero in the steady state.

Similar to equilibrium phase transitions it is 
possible in the case of absorbing phase transitions
to apply an external field~$h$ which is
conjugated to the order parameter, i.e., the field  
causes a spontaneous creation of active particles
(see for instance~\cite{HINRICHSEN_1}).
A realization of the external field 
for absorbing phase transitions with a conserved field
was recently developed in~\cite{LUEB_22} where the
external field triggers movements of inactive
particles which may be activated in this way.
At the critical density~$\rho_{\scriptscriptstyle \text c}$
the order parameter and its fluctuations scale as
$\rho_{\scriptscriptstyle \text a} \propto h^{\beta/\sigma}$ 
and 
$\Delta\rho_{\scriptscriptstyle \text a} \propto h^{-\gamma^{\prime}/\sigma}$, 
respectively.


Before we focus our attention to the scaling behavior
at the upper critical dimension~$D_{\text c}$
we briefly reconsider the scaling behavior
below and above~$D_{\text c}$.
In both cases the order parameter obeys for
all positive values of $\lambda$ the universal scaling ansatz
\begin{equation}
a_{\scriptscriptstyle \text a} \,  
\rho_{\scriptscriptstyle \text a}(\delta\rho, h) 
\; \sim \; 
\lambda^{-\beta}\, \, {\tilde R}
(a_{\scriptscriptstyle \rho}  
\delta \rho \; \lambda, a_{\scriptscriptstyle h} h \;
\lambda^{\sigma}) .
\label{eq:scal_ansatz_EqoS}
\end{equation}
The universal scaling function ${\tilde R}(x,y)$ is the
same for all systems belonging to a given universality
class whereas all non-universal system-dependent features 
(e.g.~the lattice structure, the range of interaction,
the update scheme, etc.)
are contained in the so-called non-universal metric factors 
$a_{\scriptscriptstyle \text a}$, $a_{\scriptscriptstyle
\rho}$, and $a_{\scriptscriptstyle h}$~\cite{PRIVMAN_1}.
Using the transformation 
$\lambda \to a_{\scriptscriptstyle \text a}^{-1/\beta} \lambda$
the number of metric factors can be reduced to 
$c_{\scriptscriptstyle \rho}=a_{\scriptscriptstyle \rho}
a_{\scriptscriptstyle \text a}^{-1/\beta}$
and 
$c_{\scriptscriptstyle h}=a_{\scriptscriptstyle h}
a_{\scriptscriptstyle \text a}^{-\sigma/\beta}$.
We will see that this simple reduction is not
possible at the upper critical dimension $D_{\scriptscriptstyle \text c}$.
Thus instead of this transformation we set in the following
$a_{\scriptscriptstyle \text a}=1$ for $D \neq D_{\text c}$
in order to formulate for all dimensions a 
unified universal scaling scheme.

\begin{figure}[t]
  \includegraphics[width=7.5cm,angle=0]{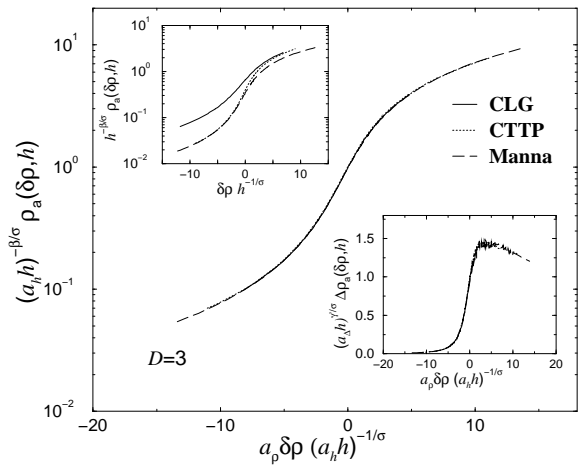}
  \caption{
    The universal scaling function of order parameter and its
    fluctuations (lower right inset) for the CLG model, the CTTP as
    well as the Manna model for $D=3$ with $\beta=0.840$
    and $\sigma=2.069$.
    The values of the non-universal metric factors are listed in
    Table\,\protect\ref{table:critical_indicees}.
    The upper left inset displays the non-universal scaling
    plots accordingly neglecting the non-universal metric factors.
    For all considered models the scaling plots contain at least
    four different curves corresponding to four different field
    values (see for \protect\cite{LUEB_22,LUEB_24} details).
   }
  \label{fig:uni_scal_3d} 
\end{figure}

The universal scaling function ${\tilde R}$ is normed
by the conditions ${\tilde R}(1,0)={\tilde R}(0,1)=1$
and the non-universal metric factors can be determined
from the amplitudes of 
$\rho_{\scriptscriptstyle \text a}(\delta \rho, h=0) \sim
(a_{\scriptscriptstyle \rho} \, \delta \rho)^{\beta}$ 
and $\rho_{\scriptscriptstyle \text a}(\delta \rho =0, h) \sim 
(a_{\scriptscriptstyle h} \, h)^{\beta / \sigma}$ .
These equations are obtained by choosing
in the scaling ansatz Eq.\,(\ref{eq:scal_ansatz_EqoS}) 
$a_{\scriptscriptstyle \rho} \delta\rho \, \lambda=1$ 
and $a_{\scriptscriptstyle h} h \, \lambda^{\sigma}=1$, respectively.
Furthermore, the choice 
$a_{\scriptscriptstyle h} h \, \lambda^{\sigma}=1$
leads to the well known scaling equation of the
order parameter
\begin{equation}
\label{eq:scal_ansatz_EqoS_collapse}
\rho_{\text{a}}(\delta \rho, h) \; \sim \; 
(a_{\scriptscriptstyle h} \, h)^{\beta/\sigma} \;
{\tilde R}(a_{\scriptscriptstyle \rho} \delta\rho 
(a_{\scriptscriptstyle h} h)^{-1/\sigma},1) .
\end{equation}
Thus plotting the rescaled order parameter 
$(a_{\scriptscriptstyle h} h)^{-\beta/\sigma}\,\rho_{\text{a}}$
as a function of the rescaled control parameter
$a_{\scriptscriptstyle \rho} \delta\rho (a_{\scriptscriptstyle h} h)^{-1/\sigma}$
the corresponding data of all systems in a given
universality class have to 
collapse onto the single curve ${\tilde R}(x,1)$.
This is shown in Fig.\,\ref{fig:uni_scal_3d} for the 
CLG model, the CTTP and the Manna model for $D=3$.
In the case that metric factors are neglected one observes
the non-universal scaling behavior where each model is
characterized by its own  scaling function (see inset of 
Fig.\,\ref{fig:uni_scal_3d}).

Similar the order parameter 
fluctuations are expected to obey the scaling
ansatz
\begin{equation}
a_{\scriptscriptstyle \Delta} \,
\Delta \rho_{\scriptscriptstyle \text a}(\delta\rho, h) 
\; = \; 
\lambda^{\gamma^{\prime}}\, \, {\tilde D}
(a_{\scriptscriptstyle \rho} \delta \rho \; \lambda, 
a_{\scriptscriptstyle h} h \, \lambda^{\sigma})  .
\label{eq:scal_ansatz_Fluc}
\end{equation}
Again the number of metric-factors can be reduced
by a simple transformation
to $d_{\scriptscriptstyle \rho}=a_{\scriptscriptstyle \rho}
a_{\scriptscriptstyle \Delta}^{1/\gamma^{\prime}}$
and 
$d_{\scriptscriptstyle h}=a_{\scriptscriptstyle h}
a_{\scriptscriptstyle \Delta}^{\sigma/\gamma^{\prime}}$.
But it is instructive to use the above 
ansatz [Eq.\,(\ref{eq:scal_ansatz_Fluc})]
since exactly one new metric factor ($a_{\scriptscriptstyle \Delta}$) 
is introduced for the fluctuations and furthermore the 
universal functions 
${\tilde R}$ and ${\tilde D}$ are characterized by 
the same metric factors.
Identical metric factors for ${\tilde R}$ and ${\tilde D}$
occur for instance naturally in equilibrium
thermodynamics where both functions can be in principle
derived from a single thermodynamic 
potential, e.g.~the free energy. 
In the case of non-equilibrium phase transitions one
can argue that both functions can be derived from a
corresponding Langevin equation.
Setting ${\tilde D}(0,1)=1$
the non-universal metric factor $a_{\scriptscriptstyle \Delta}$
can be determined by the amplitude of the divergence 
of $\Delta\rho_{\scriptscriptstyle \text a}$
similar to the order parameter.
In the inset of Fig.\,{\ref{fig:uni_scal_3d}}
we plot the rescaled fluctuations as a function
of the rescaled order parameter, i.e., the universal
scaling function ${\tilde D}(x,1)$.
Similar to the equation of state we get a good
data collapse of the corresponding data.


\begin{figure}[t]
  \includegraphics[width=7.5cm,angle=0]{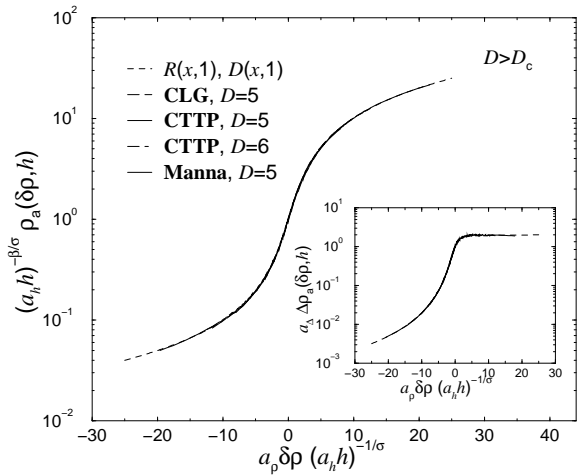}
  \caption{
    The universal scaling function of the order parameter 
    and its fluctuations (inset) above the upper critical 
    dimension~$D_{\text c}=4$ with $\beta=1$ and $\sigma=2$.
    The numerical data agree perfectly with the universal
    mean-field scaling functions ${\tilde R}(x,1)$ and
    ${\tilde D}(x,1)$ (thick dashed lines).
   }
  \label{fig:uni_scal_mf} 
\end{figure}

We consider now the scaling behavior above the 
upper critical dimension~$D_{\text c}$.
According to the renormalization group scenario
the stable fix-point of the renormalization equations
is usually the trivial fix point with classical
(mean-field) universal quantities.
Thus, in contrast to the situation below $D_{\text c}$
the critical exponents as well as the universal
scaling functions are independent of the particular
value of the dimension for $D>D_{\text c}$.
In most cases it is possible to derive these 
mean-field exponents and even the scaling functions 
exactly since correlations and fluctuations can be 
neglected above $D_{\text c}$.
The mean-field scaling behavior of the CLG model 
and the CTTP was considered in~\cite{LUEB_25}
and agrees with that of directed percolation,
i.e., the scaling functions are given 
by~\cite{MORI_1,LUEB_25}
${\tilde R}(x, y)=x/2  + [y  + (x/2)^2]^{1/2} $
and 
${\tilde D}(x , y) = {\tilde R}(x,y) 
[y  + ( x/2 )^2 ]^{-1/2}
\label{eq:uni_scal_mf_D}$.
One can easily show that $\beta=1$, $\sigma=2$, and
$\gamma^{\prime}=0$.
The latter case corresponds to a jump of the
fluctuations at the critical point which was already
observed in numerical simulations~\cite{LUEB_22,LUEB_24}.


In Fig.\,\ref{fig:uni_scal_mf} we plot the rescaled
order parameter as well as the rescaled order
parameter fluctuations for $D=5$ and $D=6$.
In all cases the numerical data are in a
perfect agreement with the mean-field scaling 
functions ${\tilde R}(x,1)$ and 
${\tilde D}(x,1)$, respectively.
Thus we clearly get the upper bound for the critical 
dimension, namely $D_{\text c}<5$.
This is a non-trivial result since a recently 
performed phenomenological field theory predicts
the too large value $D_{\text c}=6$~\cite{WIJLAND_1}.



\begin{figure}[t]
  \includegraphics[width=7.5cm,angle=0]{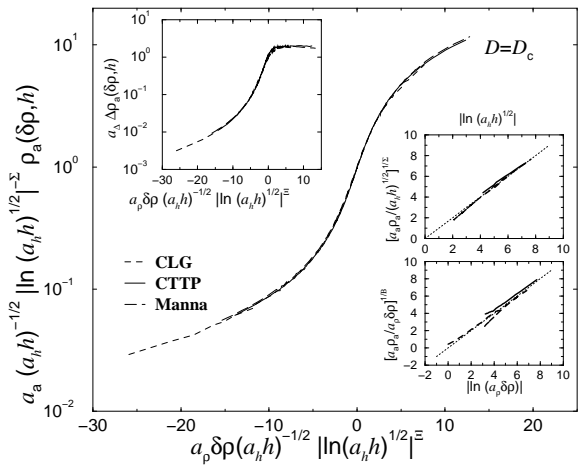}
  \caption{
    The universal scaling function at the upper critical dimension
    $D_{\text c}=4$.
    The right insets show the order parameter at the critical density 
    and for zero field, respectively.
    The order parameter is rescaled according to 
    Eqs.\,(\protect\ref{eq:uni_scal_OPzf},\protect\ref{eq:uni_scal_OPcp}).
    Approaching the transition point 
    ($h\to 0$ and $\delta\rho \to 0$)
    the data tend to the function
    $f(x)=x$ (dotted lines) as required.
   }
  \label{fig:uni_scal_4d} 
\end{figure}

We now address the question of the scaling behavior
at the upper critical dimension  $D_{\text c}=4$.
Here the scaling behavior is governed by the mean-field
exponents modified by logarithmic corrections.
For instance the order parameter obeys in 
leading order
$\rho_{\scriptscriptstyle \text a}(\delta\rho, h=0) 
\propto 
\delta \rho \, | \ln{ \delta \rho}|^{\text B}$ and
$\rho_{\scriptscriptstyle \text a} (\delta\rho=0, h) 
\propto 
\sqrt{ h}
\, | \ln{h}|^{\Sigma}$,
respectively.
The logarithmic correction exponents $\text{B}$ and 
$\Sigma$ are characteristic features of the whole
universality class similar to the usual critical exponents.
Thus it was rather surprising that recent numerical
investigations of the CLG model (${\text B}=0.24$, $\Sigma=0.45$) 
and of the CTTP (${\text B}=0.15$, $\Sigma=0.28$) 
reveals different values of the logarithmic
correction exponents~\cite{LUEB_24}.
In the following we will develop a complete scaling
scenario at the upper critical dimension which
agrees which the RG conjecture, i.e., all considered models are
characterized by the same critical exponents, 
the same logarithmic correction exponents as well as 
the same universal scaling functions.

As argued in~\cite{LUEB_22} we assume that the
universal scaling ansatz of the order parameter
obeys in leading order 
\begin{equation}
a_{\scriptscriptstyle \text a}  \, \rho_{\scriptscriptstyle \text a}(\delta\rho, h) 
\; \sim \; 
\lambda^{- \beta}\, | \ln{\lambda}|^{l} 
\; {\tilde R}
(a_{\scriptscriptstyle \rho}  
\delta \rho \; \lambda \, | \ln{\lambda}|^{b} , 
a_{\scriptscriptstyle h} h \;
\lambda^{\sigma}\, | \ln{\lambda}|^{s}) .
\label{eq:uni_scal_EqoS_dc}
\end{equation}
Thus the order parameter at zero field ($h=0$) and at the
critical density ($\delta \rho=0$) is given in leading order by
\begin{eqnarray}
\label{eq:uni_scal_OPzf}
a_{\scriptscriptstyle \text a}  \, \rho_{\scriptscriptstyle \text a}(\delta\rho, h=0) 
& \sim & 
a_{\scriptscriptstyle \rho}  
\delta \rho \, | \ln{a_{\scriptscriptstyle \rho}  \delta \rho}|^{\text B} 
\; {\tilde R}(1,0) , \\
\label{eq:uni_scal_OPcp}
a_{\scriptscriptstyle \text a}  \, \rho_{\scriptscriptstyle \text a} (\delta\rho=0, h) 
& \sim & 
\sqrt{a_{\scriptscriptstyle h}  h}
\, | \ln{\sqrt{a_{\scriptscriptstyle h} h}}|^{\Sigma} 
\; {\tilde R}(0,1)
\end{eqnarray}
with ${\textrm B}=b+l$ and $\Sigma=s/2+l$ and where
we use the mean-field values $\beta=1$ and $\sigma=2$,
respectively.
Similar to the case $D \neq D_{\text c}$ we set again
${\tilde R}(0,1)={\tilde R}(1,0)=1$.

Although the universal scaling ansatz 
[Eqs.\,(\ref{eq:uni_scal_EqoS_dc}-\ref{eq:uni_scal_OPcp})]
and the non-universal scaling ansatz (without metric
factors) are asymptotically equal, they
may lead to different results for numerically 
available data.
For instance the non-universal metric factor 
in Eq.\,(\ref{eq:uni_scal_OPzf}) 
results in the correction factor 
$|1+\ln{a_{\scriptscriptstyle \rho}/\ln{\delta\rho}}|^{\text B}$
compared to the non-universal ansatz.
This factor tends to one for $\delta\rho\to 0$
but in numerical simulations $\delta\rho$ is hardly 
smaller than $10^{-3}$
which explains why different values of 
${\text B}$ and $\Sigma$ are observed 
numerically~\cite{LUEB_24}.


\begin{table}[b]
\caption{The non-universal quantities for various dimensions.
The uncertainty of the metric factors is less than 5\%.
For greater uncertainties the corresponding data sets display
significant deviations from the presented universal 
scaling plots.}
\label{table:critical_indicees}
\begin{tabular}{lllllll}
Model & $D\;\;$ & $\rho_{\scriptscriptstyle \text c}$ &
$a_{\scriptscriptstyle \text a}$  & 
$a_{\scriptscriptstyle \rho}$     &
$a_{\scriptscriptstyle h}$  	  &
$a_{\scriptscriptstyle \Delta}$   \\
\colrule \\
CLG,  & $3$	& $0.21791\pm0.00009\;\;$ & $1$		& $0.434$	& $0.391$	& $8.881$ 
\\
CTTP  & $3$	& $0.60489\pm0.00002$ & $1$		& $0.384$	& $0.093$	& $24.51$  \\
Manna$\;$& $3$	& $0.60018\pm0.00004$ & $1$		& $0.311$	& $0.074$	& $32.24$  \\
CLG   & $4$	& $0.15705\pm0.00010$ & $4.307\;$	& $1.664\;$	& $8.021\;$	& $7.327\;$  \\
CTTP  & $4$	& $0.56705\pm0.00003$ & $0.689$	& $0.269$	& $0.047$	& $17.18$    \\
Manna & $4$	& $0.56451\pm0.00007$ & $0.690$	& $0.245$	& $0.040$	& $18.82$    \\
CLG   & $5$	& $0.12298\pm0.00015$ & $1$		& $0.329$	& $0.665$	& $8.971$  \\
CTTP  & $5$	& $0.54864\pm0.00005$ & $1$		& $0.461$	& $0.251$	& $18.73$  \\
CTTP  & $6$	& $0.53816\pm0.00007$ & $1$		& $0.421$	& $0.218$	& $157.5$  \\
Manna & $5$	& $0.54704\pm0.00009$ & $1$		& $0.870$	& $0.225$	& $20.69$ 
\\
\end{tabular}
\end{table}

According to the ansatz Eq.\,(\ref{eq:uni_scal_EqoS_dc})
the scaling behavior of the equation of state 
is given in leading order by
\begin{equation}
a_{\scriptscriptstyle \text a}  \, \rho_{\scriptscriptstyle \text a}(\delta\rho, h) 
\; \sim \; 
\sqrt{a_{\scriptscriptstyle h}  h}
\; | \ln{\sqrt{a_{\scriptscriptstyle h} h}}|^{\Sigma} 
\; {\tilde R}(x,1) 
\label{eq:uni_scal_EqoS}
\end{equation}
where the scaling argument is given in leading order by 
$ x =  a_{\scriptscriptstyle \rho} \delta\rho 
\sqrt{a_{\scriptscriptstyle h} h\,}^{-1} \,
| \ln{\sqrt{a_{\scriptscriptstyle h}  h}}|^{\Xi} $
with $\Xi=b-s/2={\text B}-\Sigma$.
Similarly we use for the 
order parameter fluctuations the ansatz
\begin{eqnarray}
& a_{\scriptscriptstyle \Delta} & \, \Delta\rho_{\scriptscriptstyle \text a}(\delta\rho, h) 
\; \sim \;  \\
& & \lambda^{\gamma^{\prime}}  \, | \ln{\lambda}|^{k} 
\; {\tilde D}
(a_{\scriptscriptstyle \rho}  
\delta \rho \; \lambda \, | \ln{\lambda}|^{b} , 
a_{\scriptscriptstyle h} h \;
\lambda^{-\sigma}\, | \ln{\lambda}|^{s}) . \nonumber
\label{eq:uni_scal_fluc_dc}
\end{eqnarray}
Using the mean-field value $\gamma^{\prime}=0$
and taking into account that the order parameter
fluctuations remain finite 
at $D_{\text c}$~\cite{LUEB_22,LUEB_24} (i.e.~$k=0$)
we get the scaling function
$a_{\scriptscriptstyle \Delta}  \, \Delta\rho_{\scriptscriptstyle \text a}(\delta\rho, h) 
\sim  {\tilde D}(x,1)$ .
The non-universal metric factor $a_{\scriptscriptstyle \Delta}$
is determined by the condition ${\tilde D}(0,1)=1$.

Thus the scaling behavior of the order parameter and 
its fluctuations at the upper critical dimension
is determined by two independent exponents (${\text B}$
and $\Sigma$) and four non-universal metric factors
($a_{\scriptscriptstyle \text a},a_{\scriptscriptstyle \rho},
a_{\scriptscriptstyle h}, a_{\scriptscriptstyle \Delta}$).
We determine these values in our analysis by the 
following conditions which are applied simultaneously:
first, both the rescaled equation of state and the rescaled
order parameter fluctuations have to collapse to the 
universal functions ${\tilde R}(x,1)$ and ${\tilde D}(x,1)$ 
for all considered models.
Second, the order parameter behavior at zero field and at the 
critical density is asymptotically given by the simple
function $f(x)=x$ if one plots 
$[a_{\scriptscriptstyle \text a} \rho_{\scriptscriptstyle \text a}(\delta\rho,0)
/a_{\scriptscriptstyle \rho} \delta\rho]^{1/{\text B}}$
vs.~$|\ln{a_{\scriptscriptstyle \rho} \delta\rho}|$
and 
$[a_{\scriptscriptstyle \text a} \rho_{\scriptscriptstyle \text a}(0,h)
/\sqrt{a_{\scriptscriptstyle h} h\,}]^{1/{\Sigma}}$
vs.~$|\ln{\sqrt{a_{\scriptscriptstyle h} h\,}}|$,
respectively.
Applying this analysis we observed that convincing results
are obtained for $\Sigma=0.35$ and ${\text B}=0.20$ 
(see Table\,\ref{table:critical_indicees} for the values of the non-universal
scaling factors).
The corresponding plots are presented in
Fig.\,\ref{fig:uni_scal_4d}.
In particular the data collapse of the equation
of state is quite sensitive for variations of the
exponents ${\text B}$ and $\Sigma$.
Thus the quality of the corresponding
data collapse could be used in
order to estimate the error-bars of the 
logarithmic correction exponents.
We obtained
in this way $\Sigma=0.35\pm 0.06$ and ${\text B}=0.20\pm0.05$.


In conclusion, the investigation of the universal scaling behavior
presents reliable results of the logarithmic correction exponents in
contrast to the non-universal scaling analysis.
Furthermore the universal scaling analysis allows to 
determine the value of $D_{\text c}$ just by checking  
whether the numerical or experimental data are in 
agreement with the usually known universal mean-field 
scaling functions.


We would like to thank A.~Hucht  
and P.\,K.~Mohanty for helpful discussions.
This work was financially supported by the 
Minerva Foundation (Max Planck Gesellschaft).

\end{document}